 \documentstyle[12pt,graphicx]{article}  
\def\beq{\begin{equation}}
\def\eeq{\end{equation}}
\def\lsim{\ ^<\llap{$_\sim$}\ }
\def\gsim{\ ^>\llap{$_\sim$}\ }
\def\r2{\sqrt 2}
\def\beq{\begin{equation}}
\def\eeq{\end{equation}}
\def\beqn{\begin{eqnarray}}
\def\eeqn{\end{eqnarray}}

\def\PL{{1-\gamma_5\over 2}}
\def\PR{{1+\gamma_5\over 2}}
\def\sinW2{\sin^2\theta_W}

\def\mz2{M_{z}^2}
\def\c2b{\cos 2\beta}

\def\m#1{{\tilde m}_#1}

\def\mc#1{{\tilde m}_{\chi^{+}_#1}}

\def\mz{M_z}

\def\cb{\cos\beta}
\def\sb{\sin\beta}

\def\Fq2{F_{2}(q^2)}
\def\sec2w{sec^2\theta_W}
\def\gmin2{(g-2)_\mu}

\catcode`\@=11 % This allows one to modify PLAIN macros.

\def\lsim{\mathrel{\mathpalette\@versim<}}
\def\gsim{\mathrel{\mathpalette\@versim>}}
\def\@versim#1#2{\vcenter{\offinterlineskip
    \ialign{$\m@th#1\hfil##\hfil$\crcr#2\crcr\sim\crcr } }}

\def\PL{Phys. Lett.}

\def\PR{Phys. Rev.}
\def\ZPHY{{Z. Phys C} }

\begin{document}
\begin{flushright}
{TIFR/TH/01-08}\\
{CERN-TH/2001-062 }\\
%\date{\today}
\end{flushright}
%-----------------------------------
%\documentstyle[preprint,aps]{revtex}
\begin{center}
{\Large\bf Constraints on Explicit CP Violation from the\\
Brookhaven Muon $g-2$ Experiment\\}
\vglue 0.5cm
{Tarek Ibrahim$^{(a)}$, Utpal Chattopadhyay$^{(b)}$ and 
Pran Nath$^{(c,d)}$
\vglue 0.2cm
{\em 
$^a$ Department of Physics, Faculty of Science, University of 
Alexandria, Egypt}\\
{\em 
$^b$ Department of Theoretical Physics, Tata Institute of Fundamental
Research,\\
Homi Bhaba Road, Mumbai 400005, India}\\
{\em $^{(c)}$ Theoretical Physics Division, CERN CH 1211, Geneva, 
Switzerland\\}
{\em $^{(d)}$Department of Physics, Northeastern University, Boston,
MA 02115, USA\footnote{Permanent address}\\} }
\end{center}
\begin{abstract}
We use the recently derived CP phase dependent analytic results
for the supersymmetric electro-weak correction to $g_{\mu}-2$ 
to constrain the explicit CP phases in softly broken supersymmetry
 using the new physics effect seen 
in the g-2 Brookhaven measurement. 
 It is shown that the
BNL data strongly constrains the CP violating phase $\theta_{\mu}$
 (the phase of the Higgs mixing parameter $\mu$) and $\xi_2$
 (the phase of the SU(2) gaugino mass $\tilde m_2$) and as much as
 $60-90\%$ of the region in the $\xi_2-\theta_{\mu}$ plane is 
 eliminated over a significant region of the MSSM parameter
space by the BNL constraint. The region of CP phases not excluded
by the BNL experiment allows for large phases and for a satisfaction
of the EDM constraints via the cancellation mechanism.
We find several models with large CP violation which satisfy the 
EDM constraint via the cancellation mechanism and 
 produce an $a_{\mu}^{SUSY}$ consistent with the new physics
 signal seen by the Brookhaven
 experiment. The sparticle spectrum of these models lies within
 reach of the planned accelerator experiments. 
 
 \end{abstract}

\section{Introduction}
Recently the Brookhaven experiment E821 made a precise determination
of the muon anomaly $a_{\mu}=(g_{\mu}-2)/2$ and found a deviation 
from the Standard Model result at the $2.6\sigma$ level. 
The experiment finds\cite{brown}
\beq
a_{\mu}^{exp}-a_{\mu}^{SM}=43(16)\times10^{-10}
\eeq
A correction to $a_{\mu}$ is expected in supersymmetric models\cite{fayet} 
and a realistic analysis of the correction to $a_{\mu}$ was given
in the supergravity unified model with gravity mediated breaking of 
supersymmetry in Refs.\cite{kosower,yuan}. Specifically in the
analysis of Ref.\cite{yuan} it was pointed out that the supersymmetric
electro-weak correction to   $a_{\mu}$ could be as large or larger
than  the Standard Model electro-weak correction\cite{fuji,marciano}. 
The fact that the supersymmetric electro-weak
contribution to $a_{\mu}$ can be large is supported by
several later works which include constraints  
of the unification of the gauge coupling constants
using the high precision LEP data, radiative breaking of the
electro-weak symmetry, experimental bounds on sparticle masses
and relic density constraints on neutralino dark matter assuming
R parity conservation\cite{lopez,chatto}. 
Soon after the release of data\cite{brown} 
on the observation of a difference
between the experimental value and the theoretical prediction
of $a_{\mu}$ in the Standard Model 
 several analyses appeared [9-17] 
both within the supersymmetric framework [9-15] as well as 
in non-supersymmetric
scenarios\cite{others1,others2}. 
Specifically the work of Ref.\cite{un} used a 
$2\sigma$ error corridor on $a_{\mu}^{susy}$ where 
$a_{\mu}^{susy}=a_{\mu}^{exp} -a_{\mu}^{SM}$
to constrain SUSY, i.e., 
 \beq
10.6\times 10^{-10} <a_{\mu}^{susy}<76.2\times 10^{-10}
\eeq
where the error corridor also includes uncertainties due to
hadronic error\cite{davier} in the theoretical predictions. The 
analysis of Ref.\cite{un}  
showed the interesting result 
that the BNL data implies 
that the sparticles have upper limits which lie within reach
of the planned accelerator experiments. Thus, for example,
one finds that within the minimal SUGRA model\cite{sugra} 
the BNL constraint implies
that the lighter chargino mass $m_{\chi_1^{+}}\leq 600$ GeV, 
$m_{\frac{1}{2}}\leq 750$ GeV and $m_0\leq 1.1$ TeV ($\tan\beta\leq 30$)
consistent with the fine tuning criteria\cite{ccn}. These mass
ranges are within reach of the LHC and a part of the parameter 
space may also be accessible at RUNII of the Tevatron. Further,
it was shown that the BNL data implies that $sign(\mu)$
is positive using the standard sign convention\cite{sugraworking}. 
This result is 
 consistent with the experimental
$b\rightarrow s+\gamma$ constraint that eliminates much of the parameter space for the
case when $\mu<0$\cite{bsgamma} . 
 It was also discussed in Ref.\cite{un}
that the effects from extra dimensions\cite{ny2} on $g_{\mu}-2$
are typically very small and do not pose a serious background 
to the supersymmetric electro-weak contribution to $g_{\mu}-2$.

The analysis carried out in Ref.\cite{un} was under the assumption 
of CP conservation where the phases of all the soft SUSY parameters
are set to zero. However, in general the soft breaking parameters
can be complex and their presence brings in new sources of CP
violation over and above the one in the CKM matrix of the Standard Model.
The normal size of the phases is O(1) which creates a problem
in that phases of this size typically lead to the electric dipole
moment for the electron and for the neutron which are in excess
of the experimental limits\cite{commins}. Some of the possible ways to avert
this disaster consist of assuming small phases\cite{ellis},
assuming that the sparticle spectrum is heavy\cite{na} and more
recently the possibility that there are cancellations\cite{in}
which allows for large phases and a not too heavy
sparticle mass spectrum.
 For the last scenario one will have then large CP violating phases which 
would affect a variety of low energy physics, such as sparticle 
 masses and decays, Higgs mixing,  proton decay, B physics
 and baryogenesis. Thus the inclusion of CP phases is an important
 new ingredient in SUSY phenomenology.  
 Now $a_{\mu}^{SUSY}$ also depends on the CP phases and thus the 
 experimental constraints on $a_{\mu}^{SUSY}$ can be translated
 into constraints on the CP phases.

 A full analysis of the effects of CP phases on 
 $g_{\mu}-2$ in the minimal N=1 supergravity  
  and in MSSM was given in Ref.\cite{ing}.
 Remarkably it is found that the CP phases strongly
 affect $a_{\mu}^{SUSY}$ in that they can change both its sign
 and its magnitude. 
 It is this fact, i.e., that the $a_{\mu}^{SUSY}$ is a very 
  sensitive function of the CP phases that leads us to
 utilize the current BNL data to constrain the CP phases.
 The phases that enter most dominantly in the $g_{\mu}-2$
  analysis are
  $\theta_{\mu}$ and $\xi_2$
  and one finds that the BNL constraint eliminates a big chunk
  of the parameter space in  the $\xi_2-\theta_{\mu}$ plane. 
  The domains allowed  and disallowed by the BNL data depend 
  sensitively on $m_0$, $m_{\frac{1}{2}}$ and $\tan\beta$
  and less sensitively on other parameters
  (Here $m_0$ is the universal soft breaking mass for the 
  scalar fields, $m_{\frac{1}{2}}$ is the universal gaugino mass,
  and $\tan\beta=<H_2>/<H_1>$ where $H_2$  gives mass to the
  up quark and $H_1$ gives mass to the down quark and the lepton).
  Often as much $60-90\%$ of the
  area in the $\xi_2-\theta_{\mu}$ plane is excluded by the
  BNL constraint. In the limit of vanishing phases the allowed region
  reduces to the constraint $\mu>0$ which was deduced in the
  earlier analysis\cite{un} under the constraint of CP conservation.  
  Of course, not all the parameter space
   allowed by the $g_{\mu}-2$ constraint is allowed by
   the EDM constraints.  However, we find that these constraints 
   can be mutually consistent. 
   Thus we give examples of models where the phases are large,
   ie., O(1), the EDM constrains are satisfied and 
    the contribution to $a_{\mu}^{susy}$ is consistent with the
    new physics signal seen in the precise BNL measurement. 
    The outline of the rest of the
   paper is as follows: In Sec.2 we give some of the basic
   formulae which enter into the $g_{\mu}-2$ analysis with
   CP violation. In Sec.3 we give a discussion of the numerical
   results obtained by imposing the constraint of Eq.(2). 
   Conclusions are given in Sec.4

\section{$g_{\mu}-2$ with CP Violating Phases}
 $a_{\mu}^{susy}$ at the one loop level arises 
from the chargino exchange and from the neutralino exchange.
The chargino contribution is typically the  dominant one 
although the neutralino contribution can become very significant
in certain regions of the parameter space. In fact the 
neutralino exchange contribution is central in determining the 
boundary of the allowed and the disallowed region in the plane of
the CP violating phases on which $a_{\mu}^{susy}$ sensitively 
depends.
 To define notation and explain the main features of the
analysis we exhibit below the CP phase dependent chargino 
contribution and the reader is referred to the Ref.\cite{ing}
for the full analytic analysis including the neutralino 
exchange contribution. The chargino mass matrix with CP 
phases is given by

\beq
M_C=\left(\matrix{|\m2|e^{i\xi_2} & \r2 m_W  \sb \cr
	\r2 m_W \cb & |\mu| e^{i\theta_{\mu}}}
            \right)
\eeq
where we have absorbed the phases of the Higgs sector by field
redefinitions.  
 In Eq.(3) $\theta_{\mu}$ is the phase of the Higgs mixing parameter
$\mu$, $\xi_2$ is  the phase of the SU(2) gaugino mass $\tilde m_2$.
 The chargino mass 
matrix can be diagonalized by the biunitary transformation
$U^* M_C V^{-1}=diag(\mc1,\mc2)$ where U and V are  unitary 
matrices. The chargino contribution is given by\cite{ing}
\beq
a^{\chi^{+}}_{\mu}=a^{1\chi^{+}}_{\mu}+a^{2\chi^{+}}_{\mu}
\eeq
where
\beq
a^{1\chi^{+
}}_{\mu}=\frac{m_{\mu}\alpha_{EM}}{4\pi\sin^2\theta_W}
\sum_{i=1}^{2}\frac{1}{M_{\chi_i^+}}Re(\kappa_{\mu} U^*_{i2}V^*_{i1})
F_3(\frac{M^2_{\tilde{\nu}}}{M^2_{\chi_i^+}}).
\eeq
and
\beq
a^{2\chi^{+
}}_{\mu}=\frac{m^2_{\mu}\alpha_{EM}}{24\pi\sin^2\theta_W}
\sum_{i=1}^{2}\frac{1}{M^2_{\chi_i^+}}
(|\kappa_{\mu} U^{*}_{i2}|^2+|V_{i1}|^2)
F_4(\frac{M^2_{\tilde{\nu}}}{M^2_{\chi_i^+}}).
\eeq
Here
$F_3(x)$=${(x-1)^{-3}}$$(3x^2-4x+1-2x^2 lnx)$, 
  $F_4(x)$=${(x-1)^{-4}}$ $ (2x^3+3x^2-6x+1-6x^2 lnx)$ and
 $\kappa_{\mu}={m_{\mu}}/{\r2 M_W \cos\beta}$.
As discussed in  Ref.\cite{ing} the entire phase 
dependence of the chargino contribution to $a_{\mu}$ 
resides in the combination $\theta_{\mu}+\xi_2$.
The neutralino contribution, however, depends on additional phases
and one can choose these additional combinations to be 
$\theta_{\mu}+\xi_1$ and $\theta_{\mu}+\alpha_{A_{0}}$
where $\xi_1$ is the phase of the U(1) gaugino mass $\tilde m_1$
and $\alpha_{A_{0}}$ is the phase of the universal parameter $A_0$
 of trilinear soft SUSY breaking term in the scalar potential.

\section{Discussion of Results}
We want to analyze the effect of the constraint of  Eq.(2) on the
CP phases on which $a_{\mu}^{SUSY}$ sensitively depends.
We shall work in the region of the parameter space where 
sparticle masses are of moderate size and CP phases are O(1).
In this region one can manufacture  an $a_{\mu}^{SUSY}$
of the size of the new physics effect indicated by the BNL 
experiment. However, since the sparticle masses are moderate
size and the phases are O(1) we need the cancellation 
mechanism to achieve consistency with the EDM constraints.
For the purpose of the analysis we shall use the parameter space 
discussed in Ref.\cite{inmssm} which consists of the 
parameters: $m_0$, $m_{1/2}$, $A_0$, $\tan\beta$, $\theta_{\mu}$,
 $\xi_1$,  $\xi_2$ and $\xi_3$ where $\xi_1$ is phase of the U(1)
 gaugino mass $\tilde m_1$, 
 and $\xi_3$ is the phase of SU(3) gaugino mass $\tilde m_3$. 
 The electro-weak sector of the model does not involve the
SUSY QCD phase
$\xi_3$ which, however, enters in the EDM analysis of the 
neutron.  The neutralino exchange contribution depends also
on  $\xi_1$ and $\alpha_{A_0}$  in addition to its
dependence on  $\theta_{\mu}$ and $\xi_2$.
However, the dependence 
of the sum of both contributions on $\xi_1$ and $\alpha_{A_0}$ 
is weak.  Thus mainly the phases
  strongly constrained by the
Brookhaven experiment are $\xi_2$ and $\theta_{\mu}$.

In Fig.1 we  display the allowed parameter space in 
the $\xi_2-\theta_{\mu}$  plane in the range $-\pi \leq \xi_2 \leq \pi$
and $-\pi \leq \theta_{\mu} \leq \pi$
for the specific input  values
of $m_0$, $m_{1/2}$, $\tan\beta$, $A_0$, $\alpha_{A_0}$ 
and $\xi_1$ as given in the caption of Fig.1. 
As discussed in Sec.2, the chargino exchange 
contribution to $a_{\mu}^{SUSY}$ is  a function only of the
 combination   $\theta_{\mu}+\xi_2$. This means that in the 
 part of the parameter space where the chargino contribution
 is dominant a value of $\theta_{\mu}+\xi_2$ allowed by the
 BNL constraint will generate a $135^0$ line in the 
 $\xi_2-\theta_{\mu}$ plane. Similarly a range of allowed values
 of $\theta_{\mu}+\xi_2$ will generate an area at $135^0$ incline
 and we see that is approximately true in Fig.1. 
 Now, of course, if the chargino
 contribution  was the sole contribution in $a_{\mu}^{SUSY}$
 Fig.1 would consist of only parallel lines at $135^0$ incline
 within the allowed range of $\theta_{\mu}+\xi_2$ consistent with
 Eq.(2). 
 However, $a_{\mu}^{SUSY}$ also contains the neutralino exchange
 contribution which is strongly dependent  on $\xi_2$ and
 $\theta_{\mu}$ individually even when $\theta_{\mu}+\xi_2$ is fixed. 
The strong dependence of the neutralino contribution on $\xi_2$
when $\theta_{\mu}+\xi_2$  is fixed is shown in Fig.2. 
 Because of this the sum of the
 chargino and the neutralino contributions does not possess 
 the simple dependence on $\theta_{\mu}$ and $\xi_2$ in the sum form.  
   Thus in Fig.1 the boundaries
 at $135^0$ are not exactly straight lines since near the boundary 
 the neutralino contribution can move 
 $a_{\mu}^{SUSY}$ in or out of the allowed range 
 admitting or eliminating that 
 point in the parameter space of the admissible set.
 Further, since the neutralino contribution violates the 
  simple dependence on $\theta_{\mu}+\xi_2$ it 
  destroys the translational invariance of $a_{\mu}^{SUSY}$ on 
   $\theta_{\mu}$
  (with $\theta_{\mu}+\xi_2$ fixed). 
We see this violation in Fig.1 from the asymmetrical endings of the
allowed region, i.e., the lower right hand and the upper left
hand of the admissible region, are not mirror reflections of each
other. In addition to $\theta_{\mu}$ and $\xi_2$, the parameters 
 $\tan\beta$, $m_{1/2}$ and $m_0$ also have a strong
effect on $a_{\mu}$. We study the effect of changes in these below.

To study the effect of the dependence on $\tan\beta$
we carry out an analysis in Fig.3 similar to that of Fig.(1) but with
$\tan\beta=5$ and with all other parameters fixed at their values in
Fig.1. Now as pointed out in Refs.\cite{lopez,chatto}
$a_{\mu}^{SUSY}$ has a strong dependence on $\tan\beta$.
As shown in the first paper of Ref.\cite{chatto} this
dependence arises from the chiral interference term in the chargino  
exchange contribution which is proportional to $\tan\beta$ for
large $\tan\beta$. Thus a reduction in the value of $\tan\beta$ 
 reduces the magnitude of $a_{\mu}^{SUSY}$ and its relative 
smallness results in a smaller range in $\theta_{\mu}+\xi_2$ around 
$\theta_{\mu}+\xi_2=0$  consistent with Eq.(2).
Further, since the magnitude of $a_{\mu}^{SUSY}$ is smaller
 and closer to the lower limit of Eq.(2) it is more sensitive
 to the neutralino exchange contributions which can move it
  out of the allowed region more easily reducing the allowed 
 region in the $\xi_2-\theta_{\mu}$ plane which is what we see  in 
  Fig.3.

Next we study the effect of changing the value of  $m_{1/2}$.
An increase in the value of $m_{1/2}$ increases the 
chargino mass and the neutralino mass which reduces 
 the magnitude of $a_{\mu}^{SUSY}$. The reduction in 
the  magnitude of $a_{\mu}^{SUSY}$ leads to a smaller
range for $\theta_{\mu}+\xi_2$  which
is what we see in Fig.4 relative to Fig.1.
Finally we look at the effect of the
variation of $m_0$ on $a_{\mu}^{SUSY}$.
Now similar to the effect of increasing $m_{1/2}$, an increase
in the value of $m_0$ decreases the overall magnitude of 
$a_{\mu}^{SUSY}$ bringing it closer to the BNL lower limit as
given by Eq.(2). As in the analysis of Figs.3 and Fig.4
the fact that the overall magnitude of 
$a_{\mu}^{SUSY}$ is smaller means that changes  in the phase
can more easily move its value out of the BNL admissible
domain thus reducing the allowed range of $\theta_{\mu}+\xi_2$.
This is what we see in Fig.5  where $m_0=400$ GeV. 
The same argument would
indicate that a further increase in the value of  $m_0$ should
further decrease the  allowed range of $\theta_{\mu}+\xi_2$
which is what we find in Fig.6   where $m_0=600$ GeV.
We note that $a_{\mu}^{SUSY}$ is more sensitive to changes in 
$m_{\frac{1}{2}}$ than in $m_0$. This was seen already in the
analysis of Ref.\cite{un} where the upper limit of $m_0$ was found to 
be significantly  larger than the upper limit on  $m_{\frac{1}{2}}$.
The above explains why the allowed area in Fig.4 is smaller than 
in Figs.5 and 6. This is so because in Fig.4 $m_{\frac{1}{2}}$ is
close to its upper limit  while is Figs.4 and 5 $m_0$ is significantly
lower than its upper  limit.
We notice also that the allowed regions in Figs. 5 and 6 
consist of complete straight lines which means that the 
neutralino contribution role here is suppressed severely
by increasing $m_0$.

Now not all the parameter space admissible by the $g_{\mu}-2$ 
constraint in Figs. 1-6 is admissible by the constraints on the
electron and on the neutron EDM. To satisfy the EDM constraints 
via the cancellation mechanism we have to utilize also the 
parameter $\xi_3$ along with $\theta_{\mu}$,
$\xi_2$ and other soft parameters.
 (We include the two loop effects
of Ref.\cite{twoloop} in the analysis.).
 In Table 1 we exhibit five points (a-e) that lie in each of the
 allowed regions of  Fig.1 and Figs. 3-6, i.e.,
  point (a) lies in the allowed region of
Fig. 1, point (b) lies in the allowed region of Fig. 3 etc,
which satisfy the EDM constraints and the corresponding
value of $a_{\mu}^{SUSY}$ lies in the BNL range of Eq.(2).
Thus Table 1 gives five models which have
large CP violating phases, their contributions to the EDM of the
electron and of the neutron lie within experimental limits using the
cancellation mechanism and they produce a SUSY contribution to 
$g_{\mu}-2$ consistent with the signal observed by the BNL experiment.
The sparticle spectrum corresponding to cases (a-e) of  Table 1 is
shown in Table 2. One finds that in all the cases the sparticle 
spectrum is low enough that some if not all of the sparticles
must become visible at the LHC, and in some cases the sparticle
spectrum is low enough to even lie within reach of
RUNII of the Tevatron. We note that using points of Table 1 one can
generate trajectories using the scaling technique given in   
Ref.\cite{scaling} where the EDM constraints are satisfied and thus one
can use this technique to produce many more models of the type discussed
above. 

\noindent
\begin{table}[h]
\begin{center}
\caption{{Cases where 
the EDM and the g-2 experiments are 
satisfied
}}
\begin{tabular}{|l|l|l|}
\hline
\hline
(case) $\xi_2$, $\theta_{\mu}$, $\xi_3$ (radian)
 & $d_e$, $d_n$ (ecm)& 
$a_{\mu}^{SUSY}$\\
\hline
\hline
(a) $-.63$,$.3$,$.37$& $-4.2\times 10^{-27}$, $-5.3\times 10^{-26}$ & $47.0 \times 10^{-10}$\\
\hline
\hline
(b)$-.85$ ,$.4$ ,$.37$
 & $4.2\times 10^{-27}$, $4.8\times 10^{-26}$ &
$10.8\times 10^{-10}$  \\
\hline
\hline
(c)$-.8$ ,$.2$ ,$1.3$
 & $4.0\times 10^{-27}$, $5.4\times 10^{-26}$ &
$12.2\times 10^{-10}$  \\
\hline
\hline
(d)$-.32$ ,$.3$ ,$-.28$
 & $-1.2\times 10^{-27}$, $3.3\times 10^{-26}$ &
$20.1\times 10^{-10}$  \\
\hline
\hline
(e)$-.5$ ,$.49$ ,$-.5$
 & $1.8\times 10^{-27}$, $-6.6\times 10^{-27}$ &
$12.7\times 10^{-10}$  \\
\hline
\hline
\end{tabular}
\end{center}
\end{table}

\noindent
\begin{table}[h]
\begin{center}
\caption{{ Sparticle masses (in GeV) for cases (a-e) in Table 1.
}}
\begin{tabular}{|l|l|l|l|l|}
\hline
\hline
(case) &$\chi_1^0$, $\chi_2^0$, $\chi_3^0$, $\chi_4^0$ & 
$\chi_1^{+}$, $\chi_2^{+}$ &
  $\tilde \mu_1$,$\tilde \mu_2$ & $\tilde u_1$, $\tilde u_2$ \\
\hline
\hline
(a)  & $98.2$, $186.9$, $389.8$, $403.5$ &$190.2$, $405.6$ & $145.0$, $209.6$&
 $628.5$,  $647.4$ \\
(b) & $97.2$, $184.0$, $408.3$, $426.6$ & $187.0$, $426.6$ & $144.6$, $209.1$
  & $628.6$, $647.5$\\
(c) & $213.8$,$421.0$, $845.8$, $852.2$ & $429.3$,$853.2$ & $232.0$, $397.2$
&$1335.3$, $1378.0$\\
 (d) & $98.1$, $186.0$, $378.4$, $393.4$ &  $189.2$, $395.3$ &
 $413.5$, $440.3$ &$738.3$, $754.4$\\
(e) & $98.3$,$187.1$,$393.7$, $407.3$ & $190.4$, $409.3$ & $609.1$, $627.6$
& $863.2$,$877.0$\\
\hline
\hline
\end{tabular}
\end{center}
\end{table}

\section{Conclusion}
In this paper we have used the new physics signal seen by the
Brookhaven g-2 experiment and the recently derived CP phase dependent
analytic results on the supersymmetric electro-weak correction
to $g_{\mu}-2$  to put limts on the explicit CP 
violating phases that arise from the soft SUSY breaking sector
of MSSM. Using a 2 $\sigma$  error corridor around the
observed effect we find that the BNL constraint excludes a large
region in the $\xi_2-\theta_{\mu}$ plane. The amount of the
region excluded  depends sensitively of $\tan\beta$,
$m_0$, and $m_{\frac{1}{2}}$ and less sensitively on the remaining
parameters. In most of the parameter space the excluded region
is as much as $60-90\%$ of the total area, i.e., the area mapped
by $-\pi\leq \xi_2\leq \pi$, $-\pi\leq\theta_{\mu}\leq \pi$.
In the limit when the phases vanish
the allowed region limits to the case $\mu>0$ deduced in
the previous analysis using CP conservation\cite{un}.
 We also show that the regions
allowed by the BNL constraint contains points where 
the cancellation mechanism operates and 
 provide examples of models
with large CP violation consistent with the experimental EDM limits 
of the electron and 
the neutron and with SUSY contributions to $a_{\mu}$
 consistent with the new physics signal seen by the BNL experiment.
These models also possess  sparticle spectra which lie 
within reach of collider experiments planned for the near
future. \\
 \noindent
This research was supported in part by the NSF grant PHY-9901057.\\

\newpage
\begin{figure}[hbt]
\begin{center}
\includegraphics[angle=270,width=15cm]{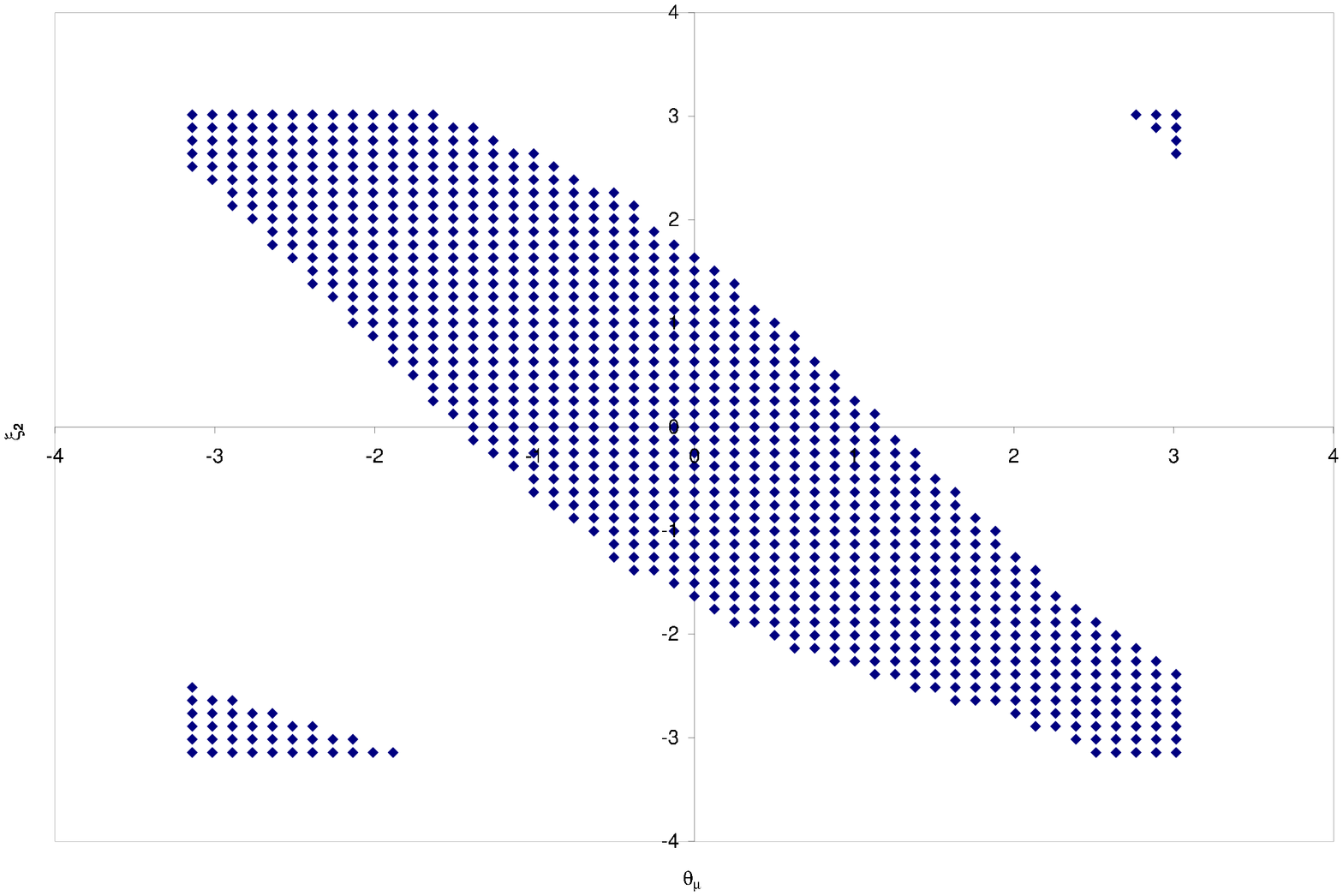}
\caption[]{A plot of the allowed region (shaded) in the 
$\xi_2-\theta_{\mu}$ plane allowed by the constraint of
Eq.(2) when $m_0=100$, $m_{\frac{1}{2}}=246$, 
$\tan\beta=20$,
$A_0$=1, $\xi_1=.3$, and $\alpha_{A_0}=.5$ where all masses
are in GeV.}
\end{center}
\label{f1}
\end{figure}

\begin{figure}[hbt]
\begin{center}
\includegraphics[angle=270,width=15cm]{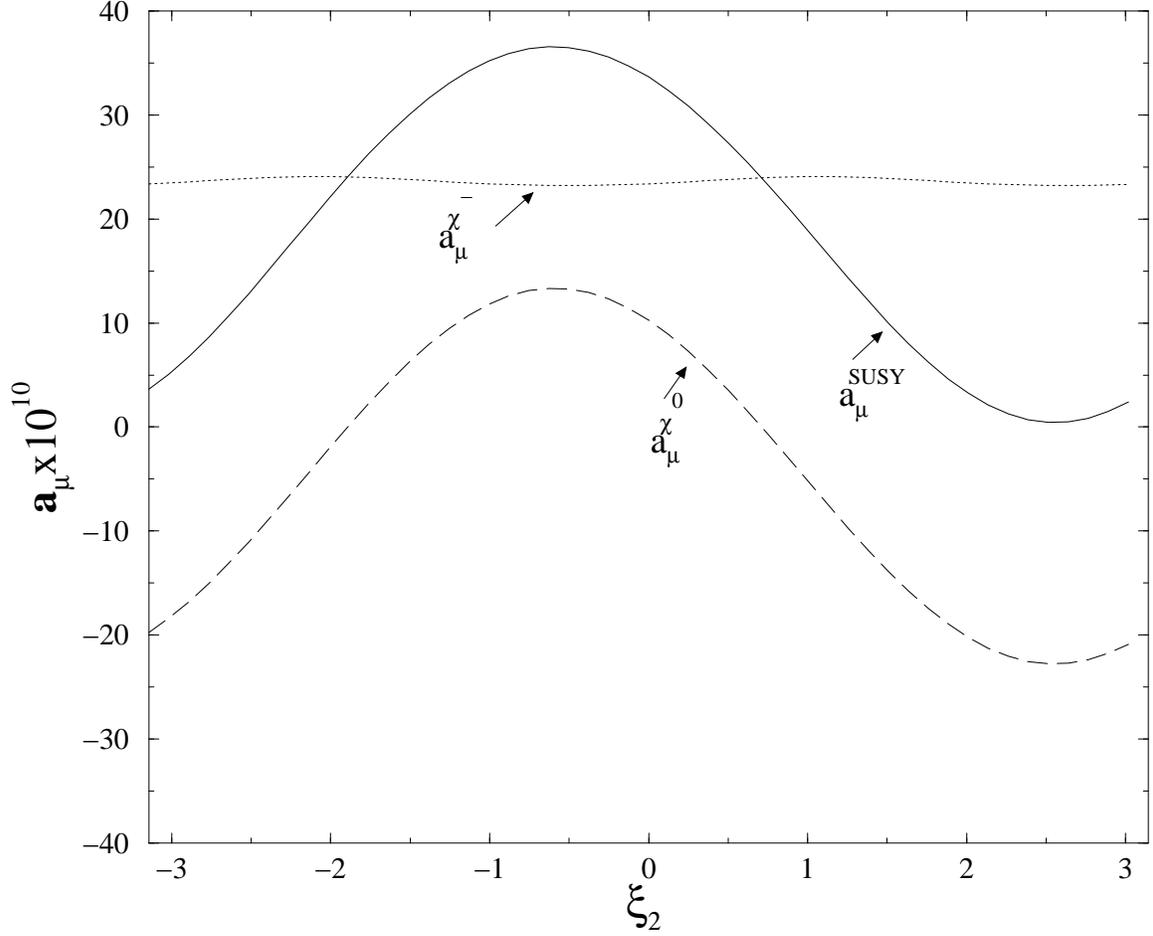}\\
\caption[]{A plot of the chargino contribution $a_{\mu}^{\chi^-}$ 
(dotted line),
neutralino contribution $a_{\mu}^{\chi^0}$
(dashed line) and the total $a_{\mu}^{SUSY}$ (solid line)
as a function of $\xi_2$ in the range $-\pi\leq \xi_2 \leq \pi$
when $\theta_{\mu}+\xi_2=-1$, $m_0=100$,
$m_{\frac{1}{2}}=246$, $\tan\beta=20$, $A_0=1$, $\xi_1=.4$, 
$\alpha_{A_0}=.5$, where all masses are in GeV. The small fluctuation of the 
chargino contribution from exact constancy is due to small 
rounding off errors in the  numerical integration program.
}
\end{center}
\label{f2}
\end{figure}

\newpage
\begin{figure}[hbt]
\begin{center}
\includegraphics[angle=270,width=16cm]{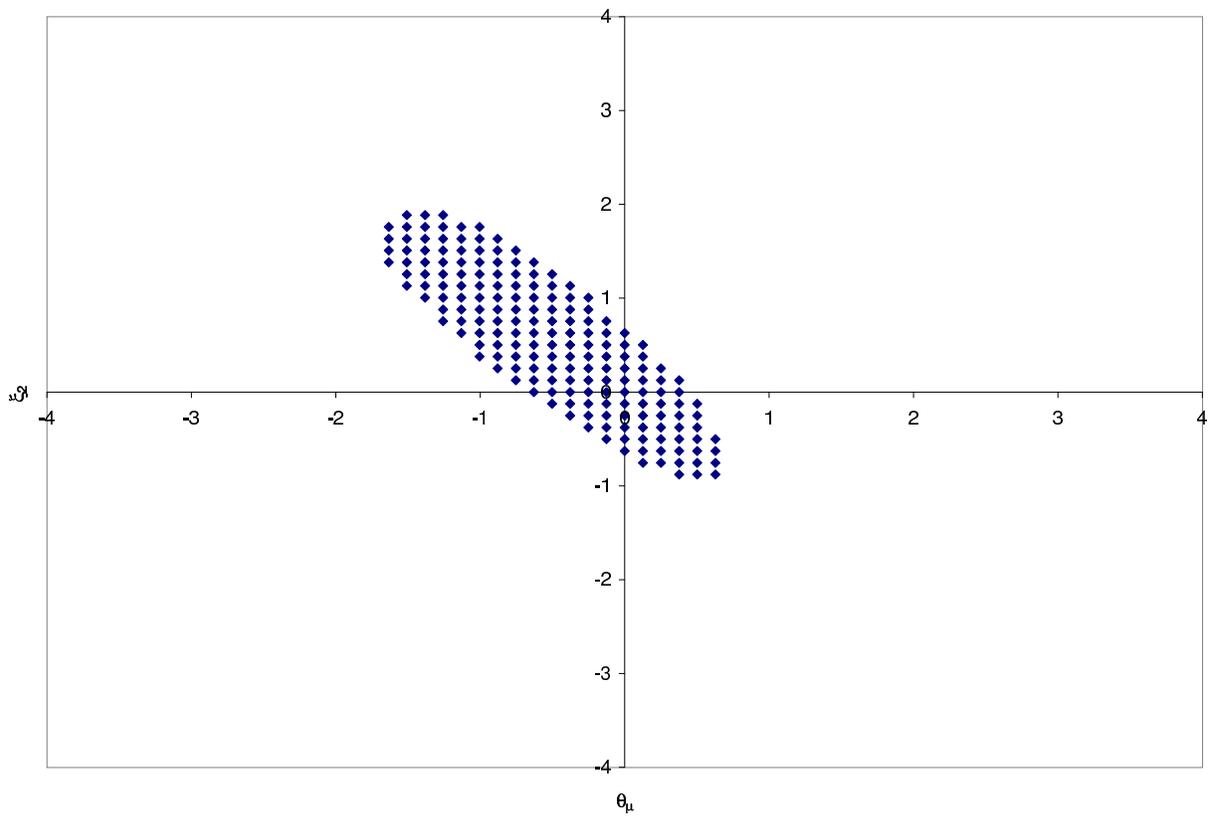}
%\vspace*{-1.0in}
\caption[]{A plot of the allowed region (shaded) in the 
$\xi_2-\theta_{\mu}$ plane allowed by the constraint of
Eq.(2) with all the same parameters as in Fig.1 except 
that $\tan\beta =5$.}
\end{center}
\label{f3}
\end{figure}

\begin{figure}[hbt]
\begin{center}
\includegraphics[angle=270,width=16cm]{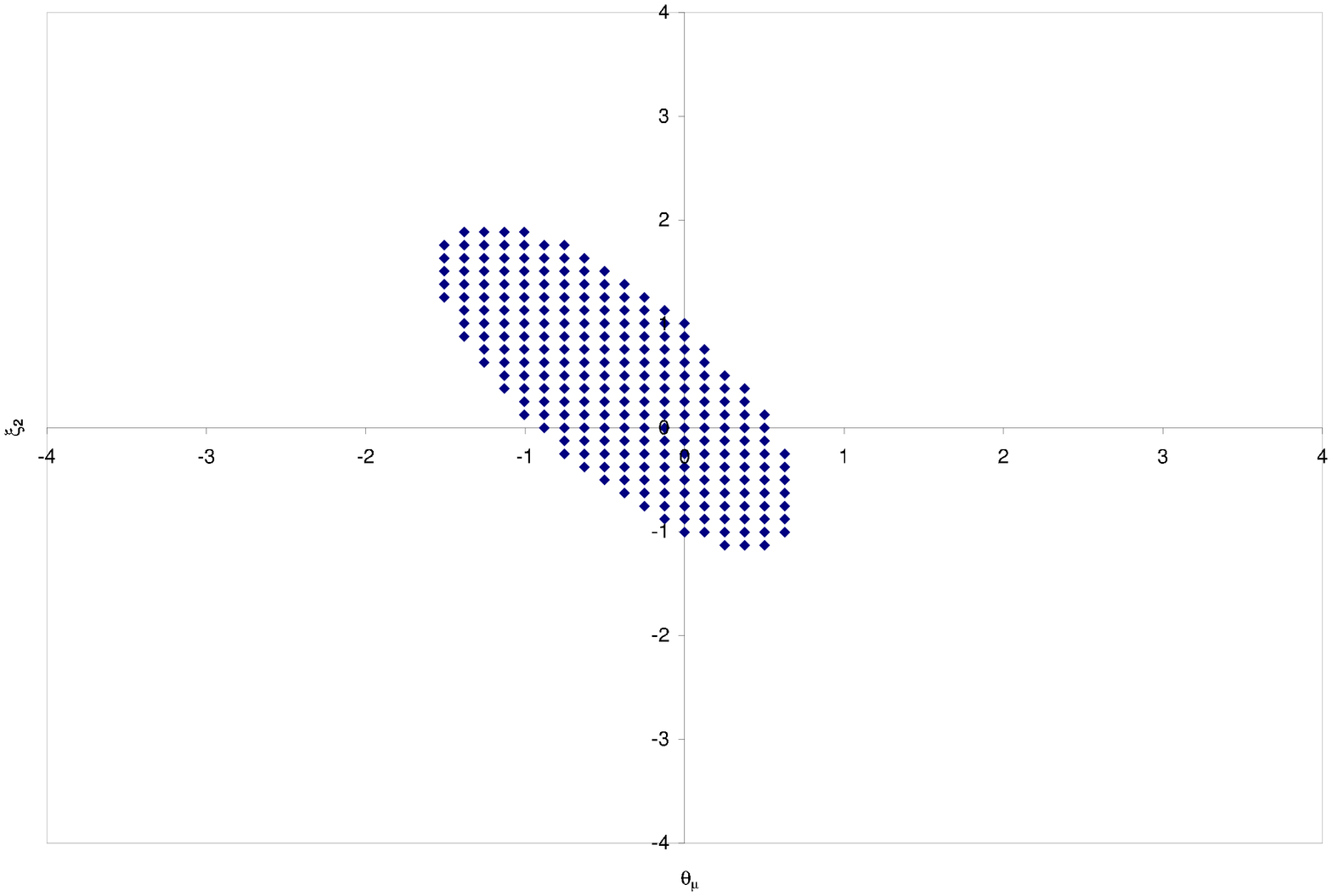}
\caption[]{
A plot of the allowed region (shaded) in the 
$\xi_2-\theta_{\mu}$ plane allowed by the constraint of
Eq.(2) with all the same parameters as in Fig.1 except 
that $m_{\frac{1}{2}}=  527$ GeV.}
\end{center}
\label{f4}
\end{figure}

\newpage
\begin{figure}[hbt]
\begin{center}
\includegraphics[angle=270,width=16cm]{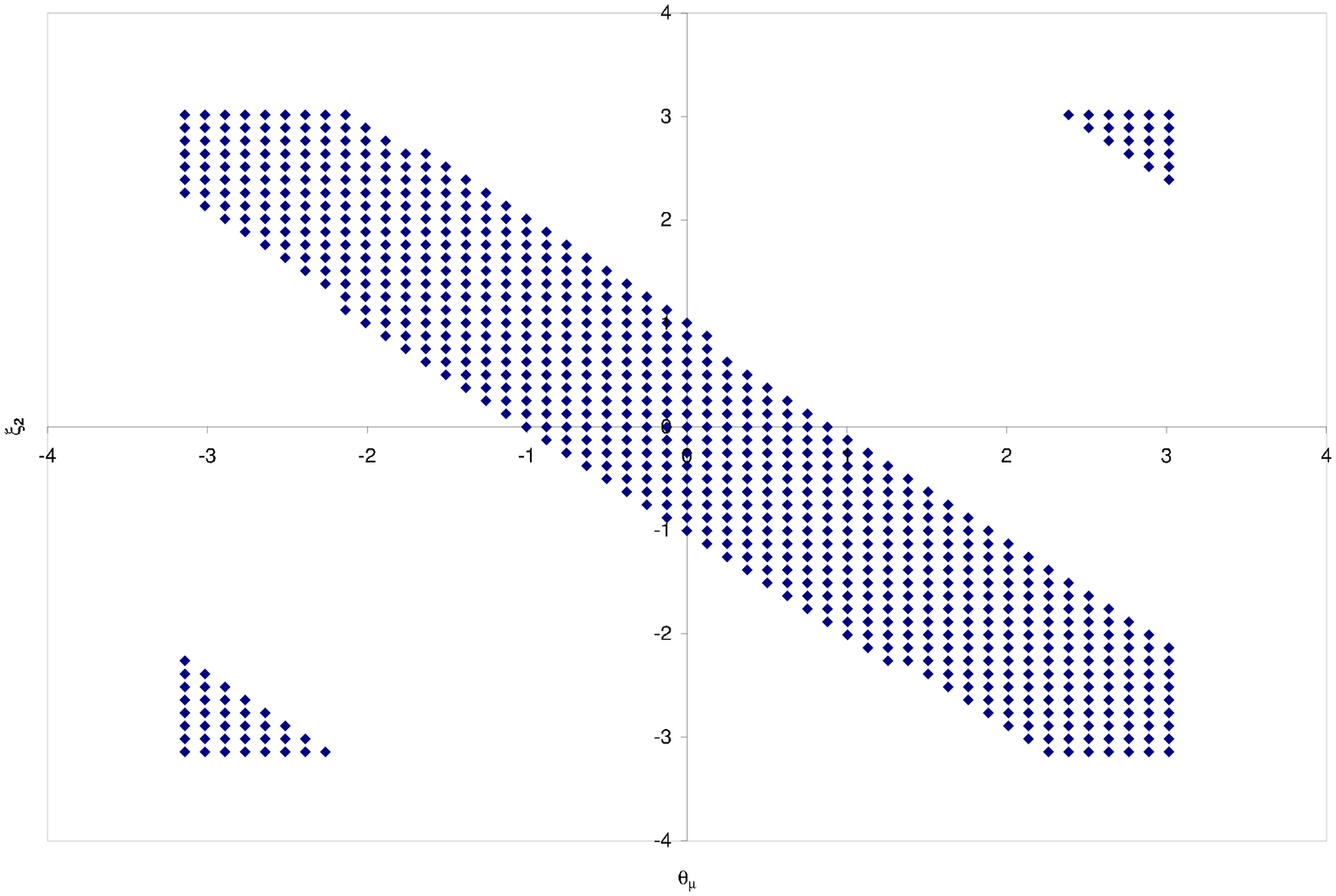}
%\vspace*{-1.0in}
\caption[]{A plot of the allowed region (shaded) in the 
$\xi_2-\theta_{\mu}$ plane allowed by the constraint of
Eq.(2) with all the same parameters as in Fig.1 except 
that $m_0= 400$ GeV.}
\end{center}
\label{f5}
\end{figure}

\begin{figure}[hbt]
\begin{center}
\includegraphics[angle=270,width=16cm]{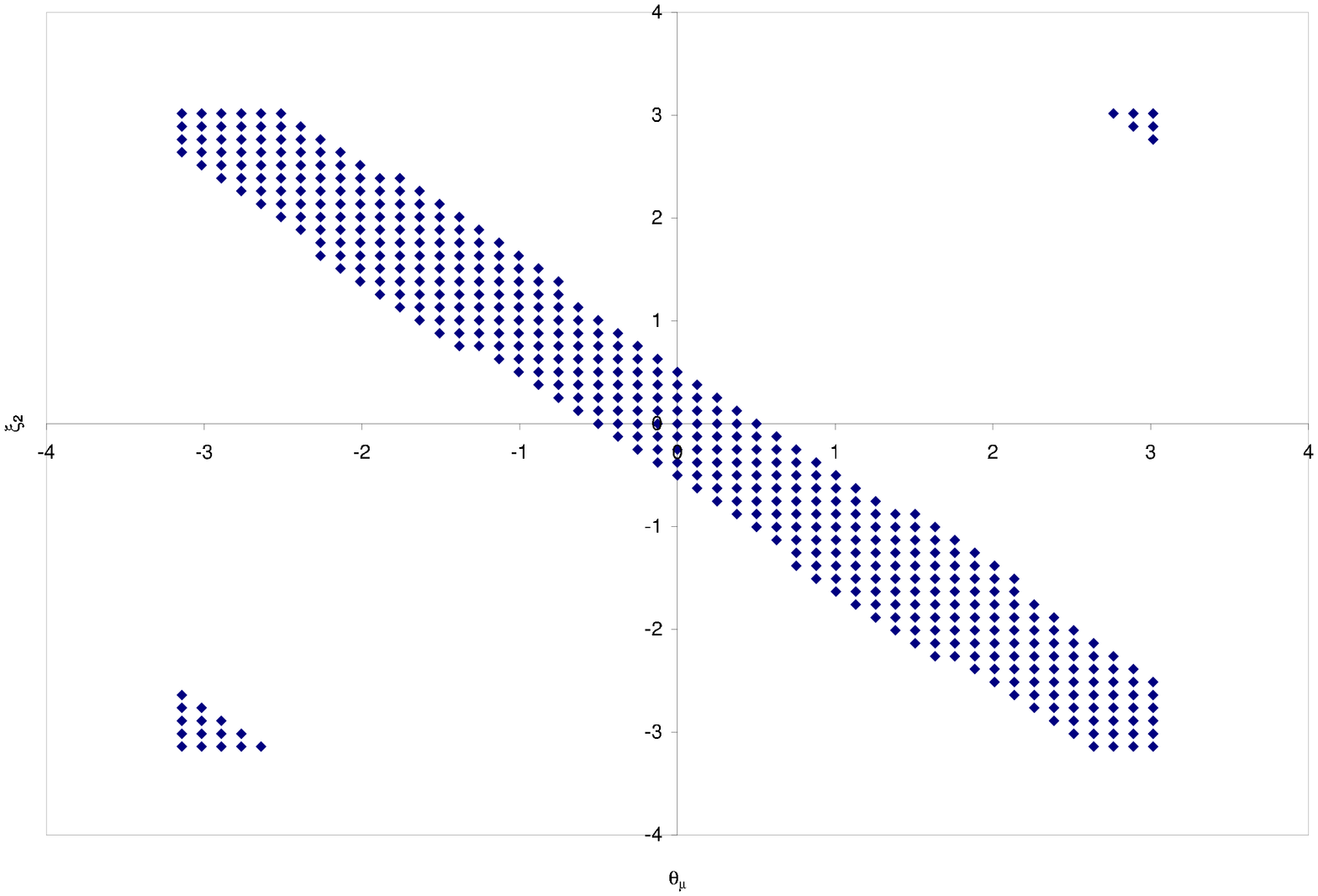}
\caption[]{A plot of the allowed region (shaded) in the 
$\xi_2-\theta_{\mu}$ plane allowed by the constraint of
Eq.(2) with all the same parameters as in Fig.1 except 
that $m_0= 600$ GeV.}
\end{center}
\label{f6}
\end{figure}

\end{document}